\documentclass[12pt]{article}
\title{\bf
Fisher Renormalization for Logarithmic Corrections}
\author{ 
{\it Ralph~Kenna$^{\,1}$,}  {\it Hsiao-Ping~Hsu$^{\,2}$}  and {\it Christian~von Ferber$^{\,1}$}\\~\\
$^1$ Applied Mathematics Research Centre,
Coventry University,\\
Coventry, CV1 5FB, England
{}\\~\\
$^2$ Institut f\"ur Physik, Johannes Gutenberg-Universit\"at Mainz\\
D-55099 Mainz, Staudinger Weg 7, Germany
 }
\begin{document}
%\pdfpagewidth  595pt
%\pdfpageheight 841pt

\maketitle
%-----------------------------------------------------------------------
                      {\Large
                      \begin{abstract}
%-----------------------------------------------------------------------
%
For continuous phase transitions characterized by power-law divergences, 
Fisher renormalization prescribes how to obtain the critical exponents for a system under constraint
from their ideal counterparts. 
In statistical mechanics, such ideal behaviour at phase transitions is frequently modified by
multiplicative logarithmic corrections.
Here, Fisher renormalization for the exponents of these logarithms is developed in a general manner.
As for the leading exponents,  Fisher renormalization at the logarithmic level is seen to be involutory
and the renormalized exponents obey the same scaling relations as their ideal analogs. 
The scheme is tested in lattice animals and the Yang-Lee problem at their upper critical dimensions,
where predictions for logarithmic corrections are made.
%
%-----------------------------------------------------------------------
                        \end{abstract} }
%-----------------------------------------------------------------------
%
  \thispagestyle{empty}
%
%***********************************************************************
%
  \newpage
%
%-----------------------------------------------------------------------
                  \pagenumbering{arabic}
%-----------------------------------------------------------------------

\setcounter{equation}{0}

There are both theoretical and practical reasons for sustained interest in thermodynamic systems subject to constraint \cite{Tr08}. 
Experimental measurements of the critical exponents characterising scaling behaviour 
at continuous phase transitions may deviate significantly from their ideal theoretical counterparts due to
the effects of such constraints. Typically, the theoretical power-law divergence of the specific heat in an ideal system
is replaced by a finite cusp in its experimental realization (often called the ``real'' system). 
Fisher \cite{Fi68} explained this phenomenon
as being due to the effect of hidden variables and established elegant relations between the exponents 
of the ideal and constrained systems (see also Ref.~\cite{ImEn73}).

Experimentally accessible examples of Fisher renormalization include 
phase transitions in constrained magnetic and fluid systems (e.g., with fixed levels of impurities) \cite{Fi68}, 
the order-disorder transition in highly compressible ammonium chloride \cite{ImEn73,Na4Cl},
the superfluid $\lambda$ phase transition present in $^3$He-$^4$He mixtures in confined films \cite{3He4He},
the critical behaviour at nematic-smectic-A transitions in liquid-crystal mixtures \cite{dePr95} and emulsions \cite{paint}
and possibly the critical behavior observed in random-field Ising systems (such as dilute antiferromagnets
in applied fields) \cite{RFIM}.

In standard notation,  the  ideal free energy is written $f_0(t,h)$, where $t$ is the reduced  temperature and $h$ is the  reduced external field.
The  family $S = \left\{{\alpha, \beta, \gamma, \delta, \nu, \eta}\right\}$ of critical exponents
characterizes the power-law  divergences of the specific-heat, magnetization, susceptibility,  correlation length and the correlation function
of the ideal system.  
The four standard scaling relations  are (see, e.g., Ref.~\cite{Fi98} and references therein)
\begin{eqnarray}
 \alpha + d \nu  & = & 2 \,,
\label{J}
\\
\alpha + 2 \beta + \gamma & = & 2  \,,
\label{R}
\\
 (\delta - 1) \beta & = & \gamma \,,
\label{G}
\\
(2 - \eta) \nu & = & \gamma \,,
\label{F}
\end{eqnarray}
where $d$ represents the dimensionality of the system.

For a system under constraint the hidden thermodynamic variable $x$ 
%may be a density of impurities (such as in the random-field Ising model \cite{CaWo86}) or ... (such as ....),  
is conjugate to a force $u$, such that 
\begin{equation}
 x(t,h,u)
 = \frac{\partial f(t,h,u)}{\partial u}
\,,
\label{218}
\end{equation}
where $f (t,h,u)$ represents the free energy of the constrained system.
The constraint is written
\begin{equation}
 x(t,h,u)
 = X(t,h,u)
\,,
\label{constraint227}
\end{equation}
where $X(t,h,u)$ is assumed to be an analytic function.
It is further assumed that the free energy of the constrained system can be written in terms of its ideal counterpart
$f_0$ as
\begin{equation}
 f(t,h,u) = f_0(t^*(t,h,u),h^*(t,h,u)) + g(t,h,u)
\,
\label{start}
\end{equation}
where $t^*$, $h^*$ and $g$ are analytic functions of their arguments \cite{Fi68}. The transition is ideal in character
if observed at fixed $u$ and the ideal free energy $f_0(t,h)$ is recovered when $u=0$. 

Under these circumstances, Fisher established that if the specific-heat exponent for the ideal system $\alpha$ 
is positive, it is renormalized in the constrained system, together with the  magnetization, susceptibility and  correlation-length 
critical exponents. If $S_X=\left\{{\alpha_X, \beta_X, \gamma_X, \delta_X, \nu_X, \eta_X}\right\}$ 
represents the family of critical exponents of the real system, it is related to $S$ via the transformation
\begin{equation}
 S_X={\cal{F}}(S)\,,
\label{Fren}
\end{equation}
where 
\begin{equation}
\alpha_X  = \frac{-\alpha}{1-\alpha}
\,,
\quad \quad
\rho_X = \frac{\rho}{1-\alpha}\,,
\end{equation}
and $\rho$ stands for any of the exponents $\beta$, $\gamma$ or $\nu$.
The exponent $\delta$ and the anomalous dimension  are not renormalized:
\begin{equation}
\delta_X = \delta  \,,
 \quad 
 {\mbox{and }}
 \quad 
\eta_X = \eta\,.
\label{Fre}
\end{equation}

These formulae have the aesthetic properties that (i) if the ideal exponents $S$ obey the scaling relations (\ref{J})--(\ref{F}) 
then the Fisher renormalized exponents $S_X$ do likewise
and  (ii) Fisher renormalization corresponds to an involutory transformation
in the sense that the ideal exponents are derived 
from the constrained ones in the same manner as the constrained from the ideal, i.e.,  applying the transformation ${\cal{F}}$ twice, gives identity: 
\begin{equation}
 S={\cal{F}}[{\cal{F}}(S)]\,.
\label{FF}
\end{equation}
Fisher's renormalization theory for power-law scaling at second-order phase transitions is well
established in statistical mechanics and its  involutory nature was studied in detail in Ref.~\cite{Sh76}. 
Dohm has discussed the possibility of Fisher renormalization  in certain circumstances where $\alpha<0$
\cite{Do74}.

The standard power-law ideal scaling behavior outlined above is frequently modified by multiplicative logarithmic corrections, 
especially in marginal scenarios \cite{We76}. 
Given the continued importance of  Fisher renormalization \cite{recent} and the ubiquity \cite{KeJo06} and 
experimental accessibility \cite{Wo00} of multiplicative logarithmic corrections in real systems, it is 
academically interesting and practically relevant \cite{Tr08} to determine their general properties under Fisher renormalization.
(For  analyses of additive corrections to scaling in constrained systems see Ref.~\cite{MrFo01}.)
Here, Fisher renormalization for logarithmic corrections is established and it is shown that 
(i)  the Fisher renormalized exponents obey the scaling relations for logarithmic corrections \cite{KeJo06} and 
(ii) Fisher renormalization is also involutory at the logarithmic level.

We consider the situation where the leading behavior of the specific heat,  magnetization, susceptibility and correlation length 
for the ideal system in the absence of an external field is
\begin{eqnarray}
 c_0 (t) & \sim & |t|^{-\alpha} |\ln{|t|}|^{\hat{\alpha}}
 \,,
 \label{C}
 \\
 \quad\quad\quad\quad m_0 (t) & \sim & |t|^{\beta} |\ln{|t|}|^{\hat{\beta}}
 \quad {\mbox{for $t<0$}}
 \,,
 \label{mt}
 \\
 \chi_0 (t) & \sim & |t|^{-\gamma} |\ln{|t|}|^{\hat{\gamma}}
 \,,
 \label{chi}
\end{eqnarray}
and 
\begin{equation}
 \xi_0 (t) \sim  |t|^{-\nu} |\ln{|t|}|^{\hat{\nu}}
 \,,
 \label{xi}
\end{equation}
respectively. 
The magnetization
at $t=0$ scales with reduced field as
\begin{equation}
 m_0(h) \sim |h|^{\frac{1}{\delta}} |\ln{|h|}|^{\hat{\delta}}
\,,
\label{mh}
\end{equation}
while, with $x$ representing position on the lattice,
the correlation function in vanishing field close to criticality takes the form
\begin{equation}
 {\cal{G}}_0 (x,t) \sim x^{-(d-2+\eta)}(\ln{x})^{\hat{\eta}} 
 D\left(
 \frac{x}{\xi(t)}
 \right)
\,,
\label{corrfun}
\end{equation}
where $D$ is a non-singular function of the ratio of length scales characterising the system.
The family of correction exponents is written ${\hat{S}} = 
\left\{{\hat{\alpha}, \hat{\beta}, 
\hat{\gamma}, 
\hat{\delta}, 
\hat{\nu}, 
\hat{\eta}, 
\hat{q}, 
}\right\}
$ in which $\hat{q}$ characterizes a possible logarithmic correction to the finite-size behaviour
of the correlation length.
In Ref.~\cite{KeJo06}, a set of scaling relations for the logarithmic-correction exponents was presented.
These scaling relations are
\begin{eqnarray}
 \hat{\alpha} + d \hat{\nu} & = & d{\hat{q}} \,, 
\label{KJJ1}\\
 \hat{\alpha} + \hat{\gamma} & = & 2\hat{\beta}  \,,
\label{KJJ2}\\
  (\delta - 1)  \hat{\beta} + \hat{\gamma} & = & \delta \hat{\delta} \,,
\label{KJJ3}\\
(2-\eta) \hat{\nu} +  \hat{\eta} & = & \hat{\gamma} \,,
\label{KJJ4}
\end{eqnarray}
where $\hat{\alpha}$ is augmented by unity in certain special circumstances
described in Ref.~\cite{KeJo06}. 
It will turn out that the Fisher renormalized logarithmic-correction exponents also obey this set of scaling relations.

Following Fisher \cite{Fi68}, the discontinuity in the magnetization at $h=0$, $t\le 0$ in the ideal system is
\begin{equation}
 \Delta m_0(t) = \lim_{ h\rightarrow 0^+ }{
                                                             \left({  m_0(t,h) - m_0(t,-h)  }\right)
                                                     }
\,,
\label{26}
\end{equation}
where
\begin{equation}
 m_0(t,h) =   \frac{ \partial f_0(t,h)}{\partial h}
\,.
\label{23}
\end{equation}
The vanishing of the discontinuity  $\Delta m_0(t) $ at $t=0$ defines this as the critical point of the ideal system.
To insure that the hidden degrees of freedom do not bias the value of 
$h$ away from zero at the transition, one assumes \cite{Fi68}
\begin{equation}
 h^*(t,h,u) = h {\cal{J}}(t,h,u)
\,,
\label{222}
\end{equation}
where ${\cal{J}}$ is also an analytic function.
From (\ref{start}), one then finds for the real system
\begin{equation}
 \Delta m(t,0,u) = \Delta m_0(t^*(t,0,u)){\cal{J}}(t,0,u)
\,,
\label{224}
\end{equation}
so that the critical point of the real system is given by that value $t_c$ of $t$
for which ${\displaystyle{t^*(t,0,u)} = 0}$. 

To accord with the scaling behaviour (\ref{C}) for the singular part of the specific heat, we assume that the
ideal free energy in the absence of field behaves for small $t$ as 
\begin{equation}
 f_0(t,0) = 
 {{\cal{A}}_0}_\pm + {{\cal{A}}_1}_\pm |t| + {{\cal{A}}_2}_\pm |t|^2 + {\cal{O}}\left({|t|^3}\right)   
 + {\cal{B}}_\pm |t|^{2-\alpha} |\ln{|t|}|^{\hat{\alpha}} \left\{{ 1 + {\cal{O}}\left({ \frac{\ln{|\ln{|t|}|}}{\ln{|t|}} }\right) }\right\}  
\,,
\label{f0}
\end{equation}
where the symbol $\pm$ corresponds to the high- and low- temperature phases. 
The internal energy is then
 \begin{equation}
 e_0(t,0) =  \frac{\partial f_0(t,0)}{\partial t} =
 A_0 + A|t|   
 + B |t|^{1-\alpha} |\ln{|t|}|^{\hat{\alpha}}  + \dots
\,,
\label{212}
\end{equation}
where $ A_0 = \pm {{\cal{A}}_1}_\pm$,
           $A = 2 {{\cal{A}}_2}_\pm$,
 and $ B   = \pm (2-\alpha) {\cal{B}}_\pm$ 
according to whether $t$ is positive or negative and where $\dots$ represents higher-order terms.

From (\ref{218}) and (\ref{start}) together with (\ref{23}) and  (\ref{222}), one obtains,
in the absence of field,
\begin{equation}
 x(t,0,u)
 = 
 e_0(t^*,0) \frac{\partial t^*(t,0,u)}{\partial u} + \frac{\partial g(t,0,u)}{\partial u} 
\,.
\label{2122}
\end{equation}
Following  Ref.~\cite{Fi68} the nonsingular function $t^*$ at $h=0$ is expanded about the real critical point 
$t=t_c$, $u=u_c$, 
\begin{equation}
 t^*(t,0,u) =  a_1 \mu + a_2 \tau + \dots \,,
\label{231}
\end{equation}
where $\mu=u-u_c$ and the reduced temperature appropriate for the contrained system is $\tau=t-t_c$.
Then, 
\begin{equation}
 \frac{\partial t^*(t,0,u)}{\partial u} = a_1 + \dots \,.
\label{squash}
\end{equation}
Now (\ref{212}) and (\ref{squash}) give for  (\ref{2122}),
\begin{equation}
 x(t,0,u) 
 =
 a_1 A_0
 + 
 a_1A  |t^*|
 + 
 a_1B {|t^*|}^{1-\alpha}|\ln{|t^*|}|^{\hat{\alpha}}
 + 
 \frac{\partial g (t,0,u)}{\partial u}
 + \dots
\,.
\label{LHS}
\end{equation}
On the other hand, expanding the constraint about the real critical point $t=t_c$, $u=u_c$, 
one has
\begin{equation}
 X(t,0,u) = X(t_c,0,u_c) + d_1 \mu+ d_2 \tau + \dots
\,
\label{230}
\end{equation}
where, from (\ref{231}), 
\begin{equation}
 \mu =  \frac{1}{a_1}t^*(t,0,u) - \frac{a_2}{a_1} \tau + \dots
\,.
\label{232}
\end{equation}

Putting (\ref{LHS}) and  (\ref{230}) into  (\ref{constraint227}), 
one obtains
\begin{equation}
 a_1^2
 \left({
A |t^*| 
+ B
 |t^*|^{1-\alpha}
 |\ln{|t^*|}|^{\hat{\alpha}}
}\right)
 = 
 d_1 t^*
 + 
 (a_1d_2 - d_1a_2) \tau
\,.
\end{equation}
If $\alpha < 0$, or $\alpha = 0$ and $\hat{\alpha} < 0$, 
the regular term dominates and $|t^*| \propto |\tau|$, leading to the absence
of Fisher renormalization.
If, on the other hand, $\alpha >0$, or $\alpha = 0$ and $\hat{\alpha} > 0$,  one obtains the central result
\begin{equation}
 |t^*| \propto |\tau|^{\frac{1}{1-\alpha}}|\ln{|\tau|}|^{-\frac{\hat{\alpha}}{1-\alpha}} 
\,.
\label{238a}
\end{equation}
The internal energy for the real system at $h=0$ is, from (\ref{start}),
\begin{equation}
 e(t,0,u) = \frac{ \partial f(t,0,u)}{\partial t}   = 
 e_0(t^*,0)\frac{\partial t^*(t,0,u) }{\partial t}
 + 
 \frac{\partial g(t,0,u)}{\partial t} 
\,.
\label{223}
\end{equation}
Now, (\ref{223}) gives for the specific heat of the real system
\begin{equation}
 c(t,0,u) \sim  |\tau|^{\frac{\alpha}{1-\alpha}}|\ln{|\tau|}|^{-\frac{\hat{\alpha}}{1-\alpha}} 
 \left\{{
   1 + {\cal{O}} \left({
 \frac{\ln{|\ln{|\tau|}|}}{\ln{|\tau|}}
                            }\right)
 }\right\}
\,.
\label{Fc}
\end{equation}
Also, from (\ref{238a}),  together with (\ref{mt})--(\ref{xi}), the scaling behaviour of the 
magnetization, susceptibility and correlation length in the constrained system is
\begin{eqnarray}
 m (t) & \sim & |\tau|^{\frac{\beta}{1-\alpha}}|\ln{|\tau|}|^{\hat{\beta}- \frac{\beta \hat{\alpha}}{1-\alpha}}
 + \dots
\,,
\label{Fm}
\\
\chi (t) & \sim & |\tau|^{\frac{\gamma}{1-\alpha}}|\ln{|\tau|}|^{\hat{\gamma}+ \frac{\gamma \hat{\alpha}}{1-\alpha}}
 + \dots
\,,
\label{Fchi}
\\
\xi (t) & \sim & |\tau|^{\frac{\nu}{1-\alpha}}|\ln{|\tau|}|^{\hat{\nu}+ \frac{\nu \hat{\alpha}}{1-\alpha}}
 + \dots
\,,
\label{Fxi}
\end{eqnarray}
respectively (where  $\dots$ again indicates higher corrections).
The leading exponents recover Fisher's result (\ref{FF}), while the exponents of the logarithms establish the following extension of the Fisher 
renormalization transformation to include the correction exponents:
\begin{equation}
( S_X, \hat{S}_X )
=
\hat{\cal{F}}[( S, \hat{S} )]
\,,
\label{Ftran}
\end{equation}
where $\hat{S}_X$ stands for the family of Fisher renormalized logarithmic correction exponents
${\hat{S}}_X = 
\{
\hat{\alpha}_X, 
\hat{\beta}_X, 
\hat{\gamma}_X, 
\hat{\delta}_X, 
\hat{\nu}_X, 
\hat{\eta}_X, 
\hat{q}_X
\}
$.
For the individual exponents of the logarithms, this results in
\begin{eqnarray}
 \hat{\alpha}_X & = & -\frac{\hat{\alpha}}{1-\alpha}
\,,
\label{Fa}
\\
 \hat{\beta}_X & = & \hat{\beta} -\frac{\beta \hat{\alpha}}{1-\alpha}
\,,
\label{Fb}
\\
 \hat{\gamma}_X & = & \hat{\gamma} +\frac{\gamma \hat{\alpha}}{1-\alpha}
\,,
\label{Fg}
\\
\hat{\nu}_X  & = &\hat{\nu} +\frac{\nu \hat{\alpha}}{1-\alpha}
\,.
\label{Fn}
\end{eqnarray}
As for the leading indices, no renormalization takes place for the logarithmic-correction exponents for the in-field magnetization 
(by construction) or the correlation function (as it is defined exactly at the critical point),
\begin{eqnarray}
 \hat{\delta}_X & = & \hat{\delta}
\,,
\label{Fd} 
\\
 \hat{\eta}_X & = & \hat{\eta} 
\,.
\label{Fe}
\end{eqnarray}
%Since the exponent $\hat{q}$ is also defined at the critical point \cite{KeJo06}, this is also not renormalized
%and $\hat{q}_X=\hat{q}$.
Note that  the case where $\alpha=0$ and $\hat{\alpha}= 1$,
which corresponds to the Ising model in two dimensions, was treated by Fisher in his original paper \cite{Fi68}
and his result for that particular case is also recovered here.

The scaling relations for logarithmic exponents ~(\ref{KJJ1})--(\ref{KJJ4})
are satisfied for the 
Fisher renormalized indices provided that $\hat{q}$ also remains unrenormalized.
Finally, it is straightforward to demonstrate that Fisher renormalization 
is also involutory at the logarithmic level: re-renormalizing $\hat{S}_X$ 
according to the prescription (\ref{Ftran})
yields the ideal counterpart $\hat{S}$.
I.e., ${\hat{\cal{F}}}$ is  its own inverse:
\begin{equation}
 (S,\hat{S}) =  {\hat{\cal{F}}}[ {\hat{\cal{F}}}[(S,\hat{S})]] .
\end{equation}

A suitable arena in which to test the above scheme is provided by lattice animals \cite{LuIs78,percolation}
in the upper critical dimensionality $D=8$, 
a system which is linked to the Yang-Lee edge problem \cite{Fi78} in $d=D-2=6$ dimensions.

The lattice animal problem deals with polyominoes or clusters of connected sites on a lattice. 
The number $A(N_b)$ of animals 
%per site 
containing $N_b$ bonds which can be embedded on a $D$-dimensional lattice
behaves as 
\begin{equation}
A(N_b) \sim K_c^{-N_b}N_b^{-\theta} (\ln{N_b})^{\hat{\theta}}
\,,
\end{equation}
where the critical fugacity $K_c$ depends on the lattice coordination number and $\theta$ and $\hat{\theta}$ are universal \cite{LuIs78}.
Lubensky and Isaacson introduced  a generating function for the number of configurations of 
branched polymers containing $N_b$ monomers  \cite{LuIs78},
\begin{equation}
f_{\rm{LA}} (K) = \sum_{N_b}{K^{N_b} A(N_b)} 
\,.
\end{equation}
This generating function is used to study the statistics of animals (clusters on a lattice)
and may be considered as the free energy of the problem
with the fugacity $K$ playing the role of the temperature in an ordinary magnetic system. 
As in the percolation problem \cite{percolation}, the free energy and other moments of $A(N_b)$ have singularities,
and these are characterised by power-law and logarithmic critical exponents. Indeed,
\begin{equation}
f_{\rm{LA}} (K) \sim |K-K_c|^{\theta-1} (\ln{|K-K_c|})^{\hat{\theta}}
\,,
\label{big}
\end{equation}
where field theory gives  $\theta=5 / 2$ and $\hat{\theta} = 1/3$  in $D=8$ dimensions \cite{LuIs78,LuIs79,AdMe88}.

The Yang-Lee problem, on the other hand, is described by a $\phi^3$ scalar field theory, 
whose upper critical dimensionality is six, with imaginary coupling \cite{Fi78}.
It  originates from the consideration of the Yang-Lee edge singularity as essentially a critical point, albeit one
with only one independent scaling field and one leading critical exponent, from which the others follow.
%The singular part of the free energy for the Yang-Lee problem in $d=6$ dimensions is
%\begin{equation}
% f_{\rm{YL}}(h) \sim h^{\sigma+1}\left({ \ln{h} }\right)^{\hat{\sigma}}
%\,,
%\label{RLf}
%\end{equation}
%where  $\sigma = 1/2$ from mean-field theory  \cite{Fi78}.
Ruiz-Lorenzo used field theory to derive the correction exponent  
for the free energy,
establishing agreement with (\ref{big}), also at the logarithmic level  \cite{RL}.
Thus  the leading critical exponents and those for the logarithmic corrections in both models coincide.
A relationship between the lattice animal problem in $D$ dimensions and the Yang-Lee singularity in $d=D-2$ dimensions
was advanced in Ref.~\cite{PaSo81} and recently rigorously established in Ref.~\cite{rig}.
%(see also Ref.~\cite{Ca03}). 

The critical exponents in  the lattice animal problem (and the Yang-Lee problem)
depend upon on the nature of the applied constraints and can be  
studied either in constant field or with constant order parameter \cite{LuIs79,LuMc81}.
Here, the order parameter conjugate to the field is the density of free ends.
If  the field  is held constant the leading exponents for the animal problem in eight dimensions
(and the Yang-Lee problem in six dimensions)  are \cite{LuIs79,Fi78}
\begin{equation}
\alpha = \frac{1}{2}\,, \quad 
\beta= \frac{1}{2}\,, \quad 
\gamma= \frac{1}{2}\,, \quad 
\delta = 2\,, \quad
  \nu=\frac{1}{4} \, \quad 
{\mbox{and}} \quad \eta=0\,.
\label{constHleading}
\end{equation}
This leading behaviour in eight dimensions has recently been confirmed numerically \cite{HsNa05}.
If the natural order parameter  of the corresponding field theory is held constant, 
then the leading exponents take on their mean-field values of  \cite{LuIs79,RL}
\begin{equation}
\alpha_X    = -1\,, \quad \beta_X  = 1\,, \quad \gamma_X = 1\,, \quad
\delta_X = 2\,, \quad \nu_X = \frac{1}{2}\,, \quad {\mbox{and}} \quad \eta_X = 0\,.
\label{constQleading}
\end{equation}
The sets of exponents (\ref{constHleading}) and (\ref{constQleading}) are related
by the standard Fisher renormalization scheme (\ref{Fren})--(\ref{Fre}).
%The set of critical exponents (\ref{constQleading}) is the most natural one for lattice animals, in the sense that 
%they are obtained from the field theory and they reduce to the usual mean-field values in high dimensionality.

Note that  both the constant order-parameter and constant field exponents violate  the hyperscaling 
relation (\ref{J}) in the lattice animal case in $D$ dimensions. 
To restore hyperscaling, the dimensionality has to be reduced to the $d=D-2$ dimensions of the Yang-Lee problem,
where all scaling relations (\ref{J})-(\ref{F}) hold \cite{PaSo81}.

Ruiz-Lorenzo determined the field-theoretical (constant order parameter) 
logarithmic corrections for the Yang-Lee problem in six dimensions as \cite{RL}
\begin{equation}
\hat{\alpha}_X    = - \frac{2}{3} \,,
\quad
\hat{\gamma}_X    = \frac{2}{3}  \,,
\quad
\hat{\nu}_X    = - \frac{5}{18}  \,.
\label{RLlog}
\end{equation}
From the scaling relations for logarithmic corrections, the remaining correction
exponents for constant order parameter are given in Ref.~\cite{KeJo06} as
\begin{equation}
\hat{\beta}_X    = 0 \,,
\quad
\hat{\delta}_X    = \frac{1}{3}  \,,
\quad
\hat{\eta}_X    =  \frac{1}{9}  \,,
\label{otherlog}
\end{equation}
with $\hat{q}_X=1/6$.

From  (\ref{Fa})-(\ref{Fg}) and (\ref{Fd}), 
Fisher renormalization for logarithmic-correction exponents now gives, for the correction exponents
in constant field,
\begin{equation}
\hat{\alpha} = \hat{\beta} =  \hat{\gamma} =  \hat{\delta}    =  \frac{1}{3} \,.
\quad
\end{equation}
These are also directly obtainable by appropriately differentiating (\ref{big}). The concurrence of these two approaches
provides an example of, and support for, the generalized scheme (\ref{Fa})-(\ref{Fg}) and (\ref{Fd}).
Finally, (\ref{Fn}) and (\ref{Fe}) yield the new predictions that the constant-field correction 
exponents for the correlation length and the correlation function in both the lattice animal and Yang-Lee problems are
$
 \hat{\nu} = \hat{\eta} = {1}/{9}
$.
These predictions now need to be independently tested.

To summarize, Fisher renormalization, which is an important and well established scheme linking the leading critical  
exponents for  ideal and constrained systems at power-law phase transitions, has been extended to deal with 
multiplicative logarithmic corrections.
%((\ref{Fa})-(\ref{Fe}), above).
The elegance of Fisher's prescription extends to the logarithmic level with the
renormalized exponents obeying the same scaling relations as the ideal ones
and renormalization scheme being involutory.
The extended renormalization scheme is supported by the examples of lattice animals and 
the Yang-Lee edge problem, where predictions are also made.
\\[4ex]
{\small{
{\emph{Note added in proof.}} 
The general results presented herein also recover those for uniaxial dipolar systems in three dimensions \cite{Ah78}
(see also \cite{Lu68}). We thank Jacques~H.H.~Perk for bringing these references to our attention.
}}

%~ \\ 
%~ \\
%%%%%%%%%%%%%%%%%%%%%%%%%%%%%%%%%%%%%%%%%%%%%%%%%%%%%%%%%%%%%%%%%%%
%\noindent
%{\bf{Acknowledgements:}}
%%%%%%%%%%%%%%%%%%%%%%%%%%%%%%%%%%%%%%%%%%%%%%%%%%%%%%%%%%%%%%%%%%%
%RK wishes to thank the Statistical Physics Group at the Universit{\'{e}}
%Henri Poincar{\'e{e}}, Nancy, France for discussions.

\bigskip
%
%%%%%%%%%%%%%%%%%%%%%%%%%%%%%%%%%%%%%%%%%%%%%%%%%%%%%%%%%%%%%%%%%%%


\begin{thebibliography}{99}
%%%%%%%%%%%%%%%%%%%%%%%%%%%%%%%%%%%%%%%%%%%%%%%%%%%%%%%%%%%%%%%%%%%

\bibitem{Tr08}
A.~Tr{\"{o}}ster, Phys. Rev. Lett. {\bf{100}}, 140602 (2008).

\bibitem{Fi68}
M.E.~Fisher, Phys. Rev. {\bf{176}}, 257 (1968).

\bibitem{ImEn73}
Y.~Imry, O.~Entin-Wohlman and D.J.~Bergman, J. Phys. C: Solid State Phys. {\bf{6}},  2846 (1973).

\bibitem{Na4Cl}
C.W.~Garland and B.B.~Weiner, Phys. Rev. B {\bf{3}}, 1634 (1971);
A.~Aharony, Phys. Rev. B {\bf{8}}, 4314  (1973).

\bibitem{3He4He}
I.D.~Lawrie and S.~Sarbach, in {\emph{Phase Transitions and Critical Phenomena}\/}, edited by C.~Domb and J.L.~Lebowitz
(Academic Press, London, 1984), Vol.9, p.1;
Y.~Deng and H.W.J.~Bl{\"{o}}te, Phys. Rev. E {\bf{70}}, 046111 (2004);
M.O.~Kimball and F.M.~Gasparini, Phys. Rev. Lett. {\bf{95}}, 165701 (2006).

\bibitem{dePr95}
{\emph{The Physics of Liquid Crystals}}, P.-G.~de~Gennes and J.~Prost, (Oxford University Press, 1995).
%J. P. Hill, B. Keimer, K. W. Evans-Lutterodt, and R. J. Birgeneau, Phys. Rev. A {\bf{40}}, 4625  (1989)

\bibitem{paint}
A.M.~Bellocq in {\emph{Handbook of Microemulsion Science and Technology}\/}, edited by P.~Kumar and K.L.~Mittal
(CRC Press, 1999) p.139.

\bibitem{RFIM}
P.-Z.~Wong, Phys. Rev. B {\bf{34}} 1864 (1986); 
R.G.~Caflisch and P.-Z.~Wong, Phys. Rev. B 34, 8160 (1986);
G.~Busiello, Phys. Stat. Sol. B {\bf{197}}, 45 (1996).

\bibitem{Fi98}
M.E.~Fisher, Rev. Mod. Phys. {\bf{70}}, 653 (1998).

\bibitem{Sh76}
D.~Shalitin, J. Phys. A {\bf{9}}, 1461 (1976).

\bibitem{Do74}
V.~Dohm, J. Phys. C: Solid State Phys.  {\bf{7}}, L174 (1974).

\bibitem{We76}
F.J.~Wegner, in {\emph{Phase Transitions and Critical Phenomena}\/}, edited by C.~Domb and M.S.~Green (Academic Press, London, 1976), Vol. VI, p.7.

\bibitem{recent}
I.M.~Mryglod, I.P.~Omelyan and R.~Folk, Phys. Rev. Lett. {\bf{86}}, 3156 (2001);
W.~Fenz, R.~Folk, I.M.~Mryglod and I.P.~Omelyan, Phys. Rev. E {\bf{75}}, 061504 (2007).

\bibitem{KeJo06} 
R.~Kenna, D.A.~Johnston, and W.~Janke, Phys. Rev. Lett. {\bf 96}, 115701 (2006);
{\emph{ibid}\/}. {\bf 97}, 155702  (2006).

\bibitem{Wo00}
W.P.~Wolf, Braz. J. Phys. {\bf{30}}, 794 (2000).

\bibitem{MrFo01}
I.M.~Mryglod and R.~Folk, Physica A {\bf{294}},  351 (2001).

\bibitem{LuIs78}
 T.C.~Lubensky and J.~Isaacson,
Phys. Rev. Lett {\bf{41}},  829 (1978); ibid {\bf{42}},  410 (1979) (erratum).

\bibitem{percolation}
D.~Stauffer and A.~Aharony, {\emph{An Introduction to Percolation Theory}\/} (Taylor \& Francis, London, 1994).

\bibitem{Fi78}
M.E.~Fisher, Phys. Rev. Lett. {\bf{40}}, 1610 (1978).

\bibitem{LuIs79}
 T.C.~Lubensky and J.~Isaacson,
Phys. Rev. A {\bf{20}},  2130 (1979).

\bibitem{AdMe88}
J.~Adler, Y.~Meir, A.B.~Harris, A.~Aharony and J.A.M.S.~Duart{\'{e}},
Phys. Rev. B {\bf{38}}, 4941 (1988).

\bibitem{RL}
J.J.~Ruiz-Lorenzo, J. Phys. A {\bf{31}}, 8773 (1998).


\bibitem{PaSo81}
G.~Parisi and N.~Sourlas, Phys. Rev. Lett. {\bf{46}},  871 (1981).

\bibitem{rig}
D.~Brydges and J.Z.~Imbrie, Annals of Mathematics   {\bf{158}},   1019 (2003);
                                             J. Stat. Phys.                    {\bf{110}},   503   (2003);
J.Z.~Imbrie,                          Annales Henri Poincar{\'{e}} {\bf{4}},  S445 (2003);
                                            J. Phys. A                            {\bf{37}},   L137 (2004).

%\bibitem{Ca03}
%J.~Cardy, cond-mat/0302495.

\bibitem{LuMc81}
T.C.~Lubensky and A.J.~McKane, J. Physique Lettres {\bf{42}},  L331 (1981).

\bibitem{HsNa05}
H.-P~Hsu, W.~Nadler and P.~Grassberger, J. Phys. A {\bf{38}},  775 (2005).

\bibitem{Ah78}
A.~Aharony, J. Magn. Magn. Mater. {\bf{7}}, 215 (1978).

\bibitem{Lu68}
A.A.~Lushnikov, Phys. Lett. A {\bf{27}}, 158 (1968);
                          Sov. Phys. JETP  {\bf{29}}, 120 (1969);
H.W.~Capel, L.W.J.~den~Ouden and J.H.H.~Perk, Physica A {\bf{95}}, 371 (1979).



\end{thebibliography}
\end{document}